%% file: main.tex
\documentclass[conference]{IEEEtran}
\IEEEoverridecommandlockouts

\usepackage{enumitem}

\usepackage{algorithm}
\usepackage{algpseudocode}

\usepackage[pdftex]{graphicx}
\graphicspath{/graphics/}
\DeclareGraphicsExtensions{.pdf}
\usepackage{tikz}

\usepackage{url}  

\begin{document}

\title{FogStore: Toward a Distributed Data Store \\for Fog Computing \thanks{This work was funded in part by DFG grant RO 1086/19-1 (PRECEPT), an NSF CPS program Award \#1446801, GTRI’s IRAD program, and a gift from Microsoft Corp.}}

\author{\IEEEauthorblockN{Ruben Mayer} 
\IEEEauthorblockA{Institute of Parallel and Distributed Systems\\
University of Stuttgart, Germany\\
Email: ruben.mayer@ipvs.uni-stuttgart.de}
\and
\IEEEauthorblockN{Harshit Gupta, Enrique Saurez, Umakishore Ramachandran}
\IEEEauthorblockA{Georgia Institute of Technology\\
Atlanta, Georgia, USA\\
Email: \{harshitg, esaurez, rama\}@gatech.edu}}


\maketitle

\begin{tikzpicture}
\begin{scope}[overlay]
\footnotesize
\node[text width=40cm] at ([yshift=-19.0cm,xshift=9cm]current page.south) {\copyright 2017 IEEE. Personal use of this material is permitted. Permission from IEEE must be obtained for all other uses,  in any current or future media, \newline including reprinting/republishing this material for advertising or promotional purposes,  creating new collective works, for resale or redistribution \newline to servers or lists, or reuse of any copyrighted component of this work in other works. \newline
This is the authors' version of the work. The definite version is published in Proceedings of 2017 IEEE Fog World Congress (FWC '17).};
\end{scope}
\end{tikzpicture}

\begin{abstract} Stateful applications and virtualized network functions (VNFs) can benefit from state externalization to increase their reliability, scalability, and inter-operability. To keep and share the externalized state, distributed data stores (DDSs) are a powerful tool allowing for the management of classical trade-offs in consistency, availability and partitioning tolerance. With the advent of Fog and Edge Computing, stateful applications and VNFs are pushed from the data centers toward the network edge. This poses new challenges on DDSs that are  tailored to a deployment in Cloud data centers. In this paper, we propose two novel design goals for DDSs that are tailored to Fog Computing: (1) Fog-aware replica placement, and (2) context-sensitive differential consistency. To realize those design goals on top of existing DDSs, we propose the FogStore system. FogStore manages the needed adaptations in replica placement and consistency management transparently, so that existing DDSs can be plugged into the system. To show the benefits of FogStore, we perform a set of evaluations using the Yahoo Cloud Serving Benchmark.
\end{abstract}


\input{content/introduction.tex}

\input{content/problem.tex}

\input{content/fogstore.tex}

\input{content/evaluation.tex}

\input{content/related.tex}
\input{content/conclusion.tex}

\IEEEtriggeratref{12}
\bibliographystyle{plain}
\bibliography{fogstore}

\end{document}

%% file: content/introduction.tex
\section{Introduction}

Large-scale distributed applications gain increasing importance in almost every aspect of life. Beyond traditional applications, via Network Function Virtualization (NFV), \emph{network functions} transform from dedicated hardware middleboxes to large-scale distributed applications. Fueled by the developments in Software-defined Networking (SDN), Virtualized Network Functions (VNFs) are increasingly prevailing. Most of the applications, SDN controllers, and VNFs are stateful \cite{Koponen:2010:ODC:1924943.1924968, Levin:2012:LCS:2342441.2342443}. Statefulness poses challenges on failure recovery and state sharing of applications \cite{7474417, 7502432}.

As an increasing trend, applications and VNFs 
are being deployed in Cloud data centers, which offer virtually unlimited resources, high availability, and reduced administration complexity.
The recent trend of Fog Computing \cite{hong2013mobile, openfog, Mayer:2017:FMS:3055601.3055614} foresees nodes with computational and storage capabilities to be placed close to the edge of the network. The network of Fog nodes builds a computational continuum between the end users' devices and the Cloud data centers. 
Different from Cloud Computing, Fog nodes are not necessarily deployed in data centers; instead, hardened routers, cellular access points, and smart home gateways also participate in the computational continuum. 
When Fog nodes are deployed at the edge of the network, they can exploit the \emph{locality} of clients, applications, and data, resulting in reduced latency and network load. The ongoing trend will push many applications and VNFs out from the central Cloud data centers toward the Fog. 

Stateful applications benefit from \emph{state externalization}, i.e., exposing their internal state to other applications, for \emph{failure recovery} and \emph{state sharing} \cite{7474417, 7502432}. Replicating and sharing state between different processes is a complex endeavor with many challenges and pitfalls, e.g., keeping consistency between the replicas and supporting concurrent read and write access. Hence, to manage externalized state, highly available \emph{distributed data stores} (DDSs) are often employed. A DDS keeps multiple replicas of the stored data records on different physical nodes. In doing so, DDSs handle the complex trade-off between consistency, availability, latency and partitioning tolerance \cite{brewer2000towards, 6127847, Pritchett:2008:BAA:1394127.1394128}. 
To benefit from Fog Computing, DDSs must be pushed from the Cloud to the Fog infrastructure. 

However, the design of current DDSs---that are designed for Cloud data centers---builds 
on assumptions that do not hold true for Fog Computing.  First, the replica placement strategies employ \emph{data center failure models}, i.e., mask failures that typically happen in a data center. For that reason, \emph{rack-aware} placement algorithms, that avoid placing multiple replicas of the same data on the same server rack, are predominant. However, such a placement is not generally applicable to Fog Computing, where we assume a more heterogeneous infrastructure that does not always provide classical server racks. Second, current DDSs do not take into account context, especially, locality of the clients; instead, all clients are provided with the \emph{globally same} read and write consistency guarantees. The assumption of uniform consistency requirements does not always hold true for Fog Computing applications. Instead, we observe, that often, the \emph{context}, e.g., the location, of clients and data determines the consistency requirements.


To tackle the shortcomings of existing DDSs with regard to Fog Computing, in this paper, we propose a Fog-enabled DDS abstraction---called \textbf{FogStore}---that implements two novel design goals: \emph{Fog-aware replica placement} and \emph{context-sensitive differential consistency}. FogStore manages the needed adaptations in replica placement and consistency management transparently, so that existing made-for-cloud DDSs can be plugged into the system. Further, we evaluate the impact of the design goals of FogStore on the overall system performance based on a data store benchmark.

%% file: content/problem.tex
\section{System Model}

Here, we introduce a system model of Fog Computing infrastructure along with an application and DDS model.

\subsection{Fog Computing Model}

\begin{figure}
    \centering
    \includegraphics[width=0.55\linewidth]{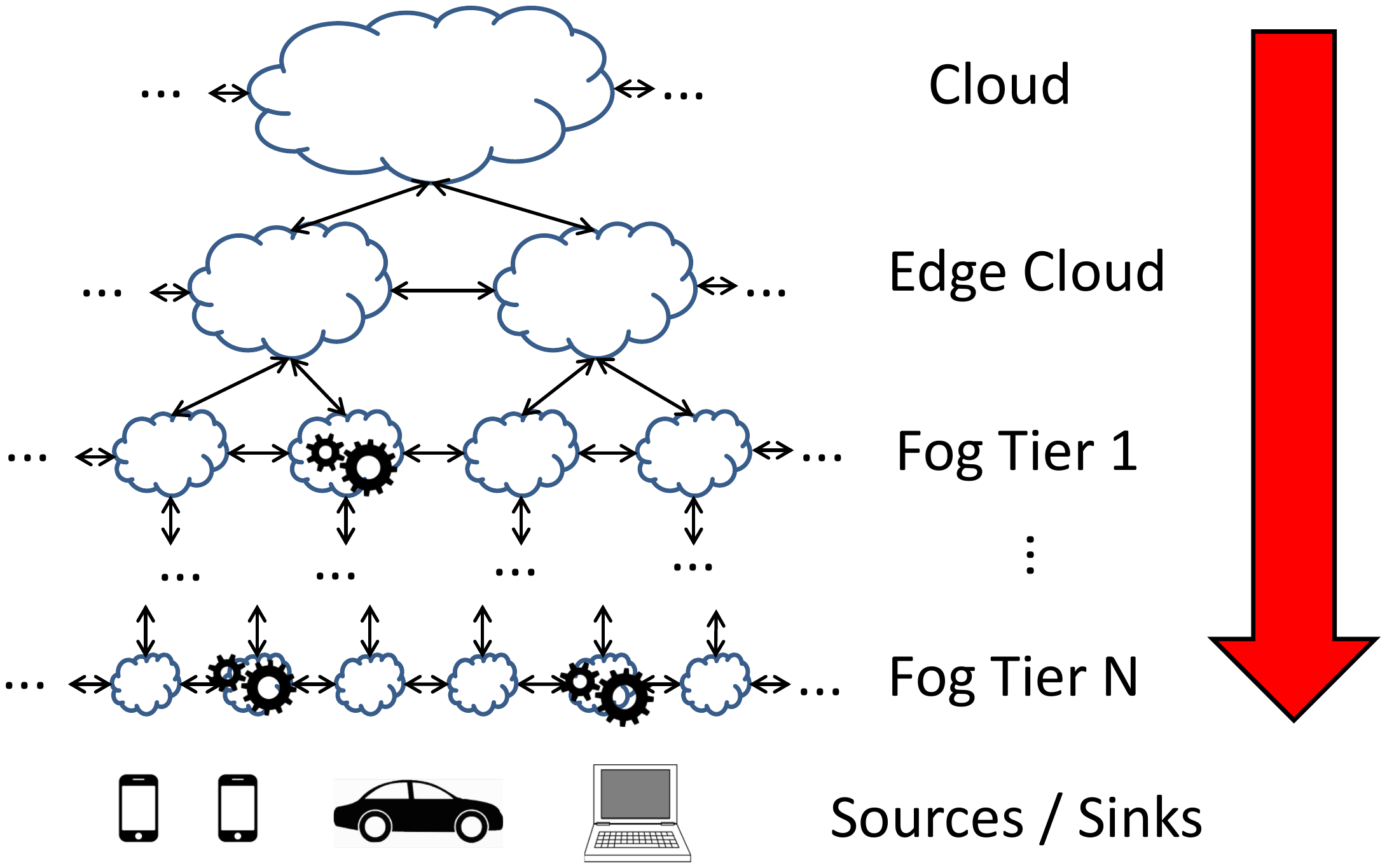}
    \vspace{-0.1cm}
    \caption{The Fog continuum provides a hierarchical network of nodes with computational and storage capabilities.}
    \label{fig:fog}
    \vspace{-0.3cm}
\end{figure}

Fog Computing describes a computational continuum between the traditional cloud data centers and the data sources and sinks \cite{openfog, Mayer:2017:FMS:3055601.3055614}. It consists of an orchestrated set of \emph{Fog nodes} that are typically hierarchically organized in multiple tiers  \cite{hong2013mobile, Saurez:2016:IDM:2933267.2933317}, as depicted in Figure \ref{fig:fog}. 
In the computational continuum of Fog Computing, a Fog node is a distinct host that provides computational and storage capabilities (denoted as the \emph{Fog platform}) and runs a software stack for deployment of applications and management and interaction between Fog nodes (denoted as the \emph{Fog software}) \cite{openfog}. The computational and storage capabilities of the Fog platforms can be heterogeneous. 
The Fog software provides an abstraction layer for the deployment of applications and services on the Fog node using virtualization technology such as containers.

In this paper, we assume that Fog nodes can fail according to the fail-recovery model, where an arbitrary number of nodes may fail and restart at any time. 
The connection between Fog nodes can be interrupted or delayed. Hence, sets of Fog nodes can become temporarily incapable of communicating with each other, i.e., temporary network partitioning is possible.

\subsection{Application and Data Store Model}
We assume \emph{stateful} applications that have an inherent \emph{locality} of their data sources and sinks, meaning that data sources and sinks are located in the same physical region. Further, we assume that the applications can expose their internal state to a DDS and read state stored in a DDS for failure recovery and state sharing between different application instances.


Multiple instances of a DDS are deployed across the Fog Computing continuum.
The data to be stored is replicated across the DDS instances---the copies of a specific data record are referred to as \emph{replicas}. In doing so, the DDS faces the problem of partitioning the data, placing the replicas of the data partitions on the available DDS instances, and keeping consistency between the replicas.

We assume that the DDS is able to support individual \emph{consistency levels} for each single read or write operation on the data. A consistency level specifies how many of the replicas of a data set need to be retrieved or updated until the operation is reported as completed to the querying client. For instance, a \emph{read consistency} level \texttt{ONE} returns a read result to the client when the data has been retrieved from one single replica, whereas a \emph{write consistency} level \texttt{QUORUM} would return a write query as successful to the client only after a majority of data replicas has been updated. Modern DDSs, such as Apache Cassandra \cite{Lakshman:2010:CDS:1773912.1773922}, allow for a fine-grained specification of the consistency level on each single (read or write) operation.

%% file: content/fogstore.tex
\begin{figure}
     \centering
     \includegraphics[width=0.7\linewidth]{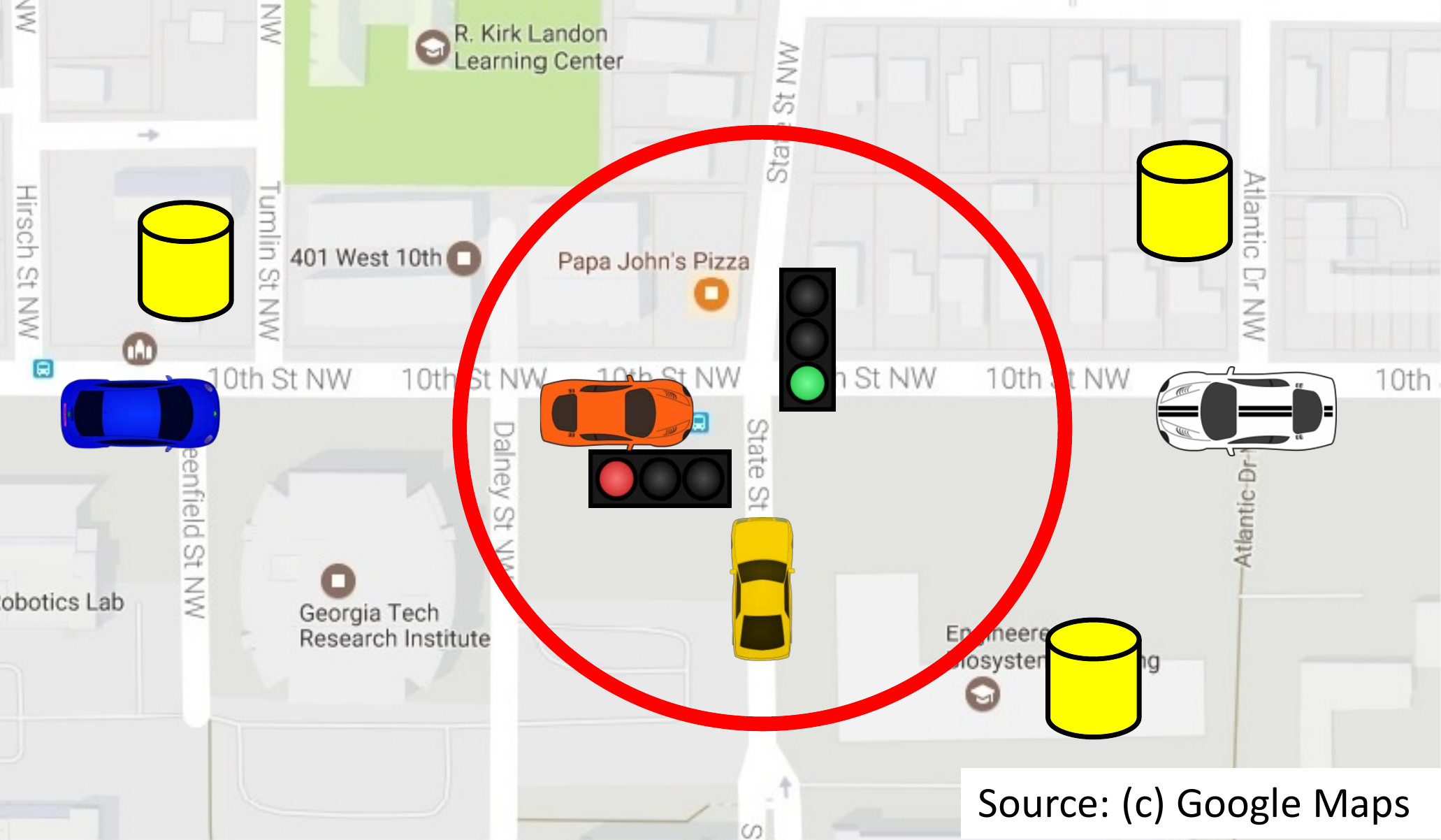}
     \vspace{-0.2cm}
     \caption{Differential consistency: only cars close to the traffic lights (within the red circle) need to read consistent data about the traffic light state.}
     \label{fig:map}
     \vspace{-0.3cm}
 \end{figure}

\section{FogStore}
\label{sec:fogstore}
To overcome the shortcomings of existing Cloud DDSs with regard to Fog Computing, we propose the FogStore system. FogStore allows for plugging in existing made-for-Cloud DDSs, extending them with Fog-aware replica placement strategies and context-sensitive differential consistency capabilities that exploit client and data locality in Fog Computing. In the following, we explain the design goals and algorithms of FogStore in more detail.



\subsection{Design Goals}
\subsubsection{Fog-aware Replica Placement}
For Cloud data centers, the placement problem has been tackled by rack-aware placement strategies that avoid to place multiple replicas of the same data in the same server rack. However, in Fog Computing, we do not assume that a placement in server rack is always appropriate or possible. The 
failures that can happen in Fog Computing are different from Cloud Computing. For instance, network partitioning can become a large problem in Fog Computing. While a Cloud data center is usually connected via redundant links to the rest of the network, in Fog Computing, whole groups of Fog nodes can be connected to the rest of the network over a single link---which might even be a wireless link. Instead of mainly focusing on server racks as the standard \emph{failure group} (i.e., a group of nodes failing together because of the same technical reason), a placement strategy for Fog Computing must also take into account the network topology and heterogeneity of the Fog nodes. 

Another common assumption of Cloud Computing is that the latency between different nodes of a Cloud data center is negligible. Typically, a customer of a Cloud service does not have full control over the placement of her virtual machines in the physical infrastructure. Contrary to that, in Fog Computing, latency between Fog nodes is a \emph{first-class citizen}; the placement of replicas in a Fog infrastructure plays a major role in the DDS performance, as also our evaluations in Section \ref{sec:evaluations} confirm. A Fog-enabled replica placement strategy should optimize the placement for achieving minimal latency in between the replicas and between the replicas and the data sources and sinks (i.e., the clients).


\begin{figure}
    \centering
	\vspace{0.1cm}    
    \includegraphics[width=0.7\linewidth]{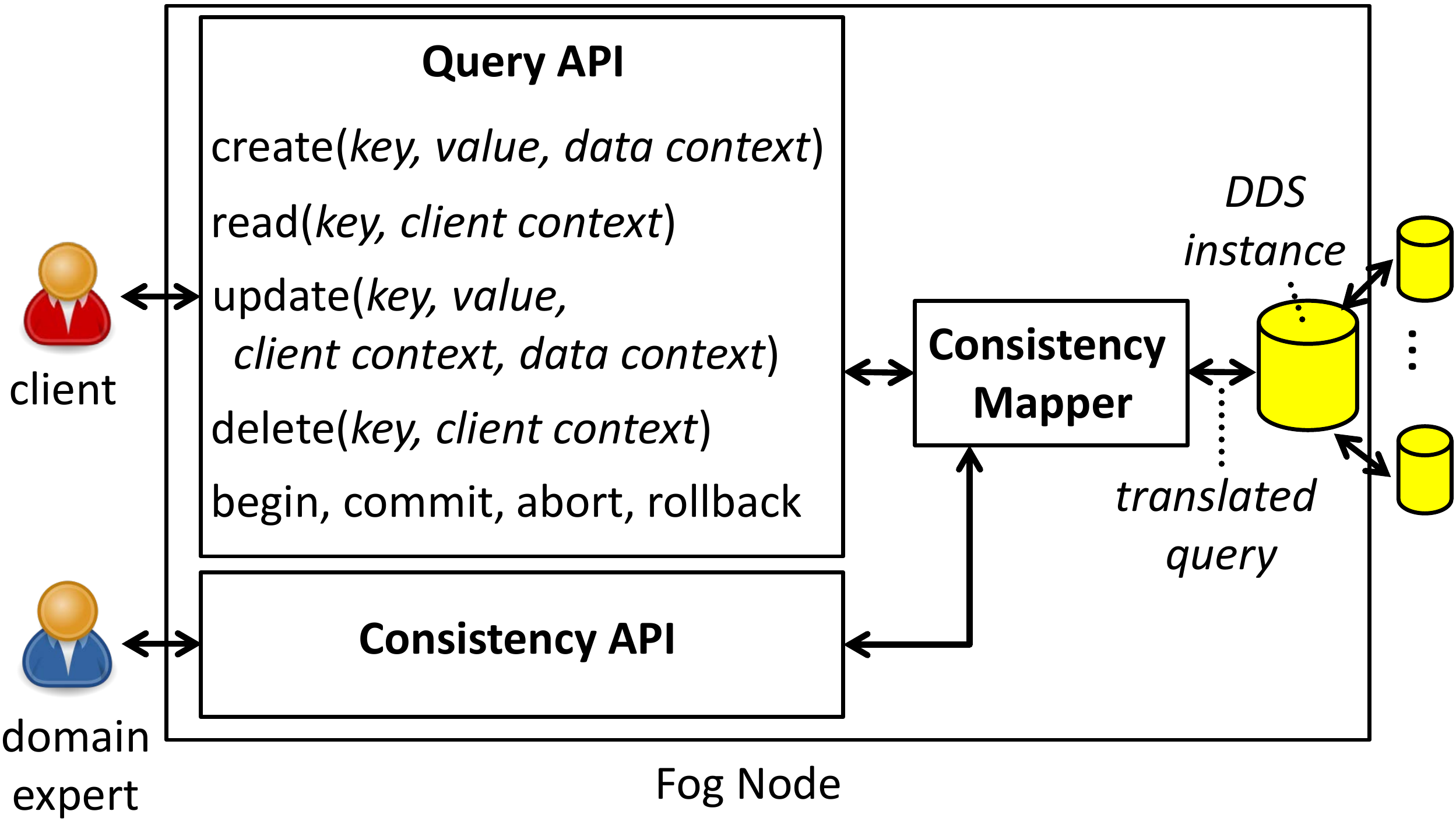}
    \vspace{-0.2cm}
    \caption{FogStore architecture.}
    \label{fig:fogstore}
    \vspace{-0.4cm}
\end{figure}


\subsubsection{Context-Sensitive Differential Consistency}
\label{sec:diffcon}

In Cloud Computing, a common assumption is that clients of a DDS are geo-distributed, concurrently accessing the same data from different locations around the globe. Hence, each client gets provided with the same consistency guarantees on its (write and read) operations on the DDS.
Contrary to that, in Fog Computing, different clients accessing the distributed data store often have an individual \emph{context}. This context can influence their requirements on consistency. FogStore allows to exploit context, in particular location, by providing a mapping from a client's context to the consistency level of the client's query. To illustrate this, we discuss an example of differential consistency, using location context, in the following.

In a situation-aware application, multiple autonomous cars read the state of a traffic light on a specific road junction (cf. Figure \ref{fig:map}). The cars need strong read consistency when they are close to the junction (i.e., within the red circle): No two cars should read contradicting traffic light status from different DDS nodes. This means that each car has to read the status from all nodes, and use the freshest one to make its decision. However, cars further away from the junction might use the traffic light status for non-critical operations such as to adjust their speed in order to save energy or to update the estimated arrival time in their navigation system. Two different cars further away from a given traffic light may read contradicting traffic light status without inducing safety risks. Hence, those cars can read from one local node, making the response faster and reducing the load on the Fog nodes.

In the following, we describe how the architecture of FogStore supports the proposed Fog-aware design goals.
 
\subsection{Architecture and Algorithms}
FogStore is an extension to existing DDSs that allows for their seamless integration into a Fog Computing environment. Each instance of the DDS, deployed on a Fog node, is plugged into an instance of FogStore. The FogStore instance receives the clients' queries and translates them to DDS queries, implementing Fog-aware replica placement and context-sensitive differential consistency. FogStore consists of three main components, as depicted in Figure \ref{fig:fogstore}: (1) A Query API to receive the create, read, update, delete and transactional queries from the DDS clients, enriched with context information. (2) A Consistency API to receive specifications of \emph{consistency regions} from a domain expert who knows how client context influences the consistency requirements. (3) A Consistency Mapper that, based on the specifications in the Consistency API, automatically maps an incoming query to an appropriate consistency level and issues a corresponding query in the plugged-in DDS. Replica placement is also handled in this component.


\begin{algorithm}[bp]
\footnotesize
\caption{FogStore Consistency Mapper}
\label{alg:consistency}
\begin{algorithmic}[1]
	\State DDS localDDS \Comment{links to closest DDS node}
    \Procedure{mapAndExecuteQuery}{query, key, client\_ctx}
    \State consistencyRegion $\gets$ \\ \qquad \quad ConsistencyAPI.\texttt{getRegion}(key, client\_ctx)
    \State consistencyLevel $\gets$ \\ \qquad \quad ConsistencyAPI.\texttt{getLevel}(consistencyRegion)
    \State translatedQuery $\gets$ 
    \texttt{translate}(localDDS.type, query, consistencyLevel)
    \State localDDS.\texttt{execute}(translatedQuery)
\EndProcedure
\end{algorithmic}
\end{algorithm}

\begin{algorithm}[bp]
\footnotesize
\caption{FogStore Replica Placement Algorithm}
\label{alg:placement}
\begin{algorithmic}[1]
    \Procedure{placement}{data\_key, data\_location, failure\_groups, fog\_topology, replication\_factor}
    \State closestFogNode $\gets$ \texttt{findClosest(}fog\_topology, data\_location\texttt{)}
    \State replicaNodes.\texttt{add(}closestFogNode\texttt{)}
    \While{replicaNodes.size $<$ replication\_factor}{}
        \State find nearest neighbor of closestFogNode that is not in same failure group
        \State add it to the replicaNodes
    \EndWhile
    \State \texttt{setDataStoreMapping(}key, replicaNodes\texttt{)}
\EndProcedure
\end{algorithmic}
\end{algorithm}

\subsubsection{Query API}

Queries are sent by the clients to the Query API of FogStore. Based on the consistency levels specified in the Consistency API, the queries are translated by the Consistency Mapper and forwarded to the underlying DDS. 

\texttt{create(key, value, data context)} -- creates a new key and value. Replicas of the data record are placed on the Fog nodes by the Consistency Mapper according to the \emph{placement algorithm}. 

 \texttt{read(key, client context)} -- retrieves the value associated with the key from the DDS. 

 \texttt{update(key, value, client context, data context)} -- updates the given key with a new value. 
By setting the field \texttt{data context} in the query, the data context can be updated, e.g., when the data source has moved to a different location. The updated data context is taken into account by the Consistency Mapper in all future queries.

 \texttt{delete(key, client context)} -- deletes the key from the distributed data store. 
 
\texttt{tx begin, commit, abort, rollback} -- allows for integrating transactional support.

\subsubsection{Consistency API}

In the Consistency API, a domain expert specifies the mapping between data context, client context, and consistency level of the different kind of queries. For location context, which is a major class of context in Fog Computing, the Consistency API provides an easy-to-use abstraction, called \emph{consistency regions}. Note, that beyond location context, other types of (application-specific) context can be extended to the user's needs. In the following, we describe the concept of consistency regions in more detail.

Consistency regions provide information about the consistency levels needed in specific geographical regions around the data location. Those specifications are usually application-specific, i.e., each application using FogStore can have individual consistency regions specified. 
For instance, in the traffic application described in Section \ref{sec:diffcon}, the consistency regions would be specified as: ``Read from \texttt{ALL} replicas when client is located around 500 meters from the data location (i.e., the traffic light location), else read from \texttt{ONE} replica''. 

\subsubsection{Consistency Mapper} The Consistency Mapper component uses the consistency regions specified in the Consistency API in order to translate client queries to the Query API into queries to the connected DDS. In doing so, an adapter to the connected DDS has to be provided, such that the consistency levels specified in the Consistency API can be translated according to the query language and consistency capabilities of the DDS. 

\paragraph{Consistency Mapping Algorithm}
Algorithm \ref{alg:consistency} lists the consistency mapping algorithm. First, the consistency region is determined based on the client's context (line 3). Then, the consistency level of that region is determined (line 5). Finally, the query and its consistency level are translated into the appropriate query language of the underlying DDS and executed (lines 7--8).

\paragraph{Replica Placement Algorithm}
Whenever a new data record is inserted into the DDS (\texttt{create} is called in the Query API), the record and its replicas will be placed on the DDS instances according to a placement algorithm. 

The Fog-aware replica placement algorithm applied in FogStore is centered around the definition of \emph{failure groups}.
A failure group is a group of nodes that will fail for the same technical reason (e.g., a local power outage). This can be a group of nodes on a rack, but also a group of nodes that would be disconnected or partitioned because they access the Fog continuum via the same physical link. 
The concrete setup of failure groups is configured by a domain expert who knows the physical conditions of the Fog continuum. In FogStore, the domain expert can group arbitrary Fog nodes in failure groups and provide them to the placement algorithm.

Algorithm \ref{alg:placement} lists the replica placement algorithm. The algorithm takes into account the location of the data, the failure groups and the Fog topology, and the required replication factor (line 1). It searches for the closest Fog node to the data in the Fog topology (line 2) to place the first replica (line 3). The further replicas are placed on the closest nodes of the first replica that are not in the same failure group (lines 4--7). When all replicas are placed, the mapping of replicas to Fog nodes is enforced on the data store (line 8). The distance between Fog nodes can be determined via network coordinates~\cite{vivaldi}.



%% file: content/evaluation.tex
\begin{figure*}
	\vspace{-0.25cm}
    \centering
    \includegraphics[width=0.75\linewidth]{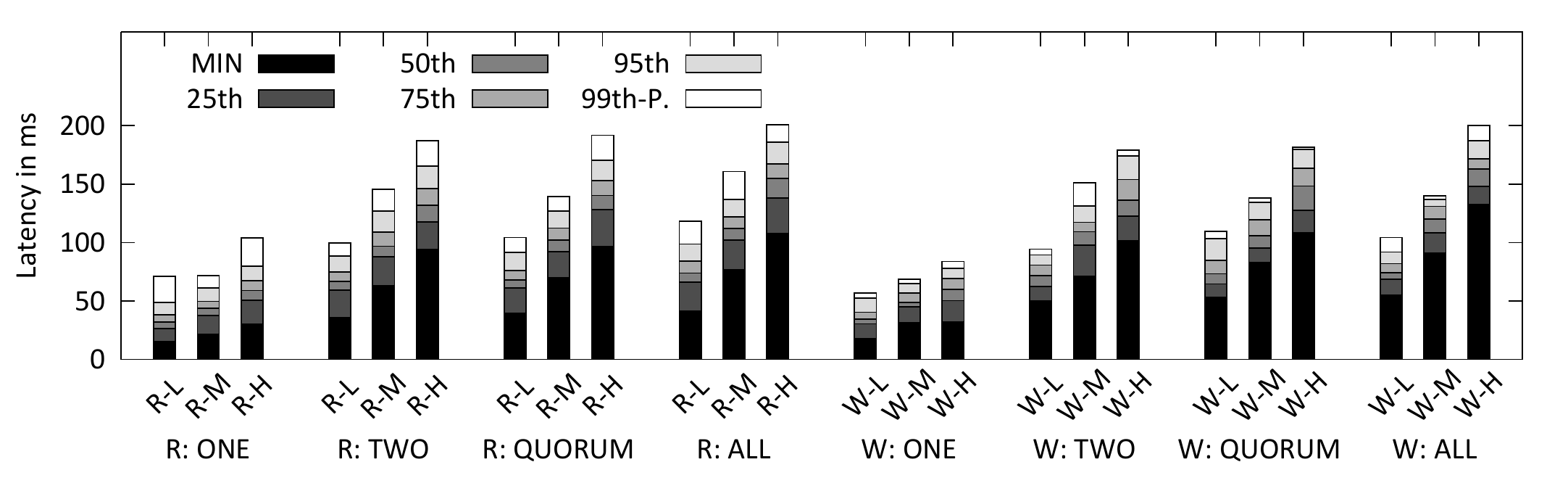}
    \vspace{-0.5cm}
    \caption{\underline{R}ead and \underline{W}rite latency on Cassandra for different consistency levels at \underline{L}ow, \underline{M}edium and \underline{H}igh network latency.}
    \label{fig:latency}
    \vspace{-0.4cm}
\end{figure*}

\section{Evaluations}
\label{sec:evaluations}

We perform micro benchmarks that indicate how latency between replicas and different consistency levels influence the read and write latency in a distributed data store.  We hasten to add that these are preliminary results simply to illustrate the concept of FogStore and the impact of its design goals.

\subsection{Evaluation Setup}
We deploy the MaxiNet network emulator \cite{6857078} on a 64 vCPU virtual machine with 128 GB of RAM and Ubuntu 16.04.2 LTS, using the OpenStack platform. The underlying hardware is a Supermicro X8OBN server with 8 x 10 Core Xeon E7 8870@2.2GHz CPUs and 1 TB RAM.

In general, the Fog nodes at a given level of the network hierarchy are assumed to form a peer-to-peer network with sufficient redundancy in connectivity to ward off link failures, node isolation, and network partitioning.
However, for the purposes of carrying out controlled experiments, 
the Fog Computing environment we deploy for our experimental study consists of 6 Fog nodes in a star topology with a switch in the middle. This topology reflects 6 different Fog nodes deployed at different locations at the network edge, which are interconnected via a backbone network represented by the central switch. We decided to use this simplification of real network topologies, because it allows for a straight-forward parameterization to simulate different physical distances between Fog nodes. 
The first 5 Fog nodes host instances of Apache Cassandra 3.10; the 6th Fog node hosts a Yahoo Cloud Serving Benchmark (YCSB) \cite{Cooper:2010:BCS:1807128.1807152} adapter, serving as the ``client'' that poses queries on the DDS. We use the YCSB-D (``read latest'') core workload that fits well with the traffic scenario from Section \ref{sec:diffcon}. All evaluation runs employ a replication factor of 5, i.e., each data set is replicated on each Cassandra node.

In the evaluations, we use 3 different \emph{network latency settings}, reflecting different placement qualities. In the \emph{low} network latency setting, the latency between the switch and Cassandra nodes is 4, 5, 6, 7 and 8 ms.  In the \emph{medium} network latency setting, the latency between the switch and Cassandra nodes is 8, 10, 12, 14 and 16 ms.  In the \emph{high} network latency setting, the latency between the switch and Cassandra nodes is 12, 15, 18, 21 and 24 ms. The latency between YCSB node and switch is 1 ms in all settings.

In each of the network settings, we run the YCSB-D workload with different consistency levels for read and write consistency. (1) Consistency level \texttt{ONE}: read/write from one replica. (2) \texttt{TWO}: read/write from two replicas. (3) \texttt{QUORUM}: read/write from a majority (i.e., three) of replicas. (4) \texttt{ALL}: read/write from all five replicas. 

\subsection{Results and Discussion}
The results of the benchmarks are depicted in Figure \ref{fig:latency}. The stacked bars represent the percentiles of latency measured per operation: minimal latency, 25th, 50th, 75th, 95th and 99th percentile. We divided the latency measurements into read latency and write latency. When measuring read latency, the consistency level for writes was fixed to \texttt{ONE}; when measuring write latency, the consistency level for reads was fixed to \texttt{ONE}.

There are two main results. First, the biggest difference in latency is between consistency level \texttt{ONE} and any other level. The latency difference between consistency level \texttt{ONE} and \texttt{TWO} is much higher than the difference between \texttt{TWO} and \texttt{QUORUM} or \texttt{ALL}. Further, consistency level \texttt{ONE} does not suffer from high latency penalties when the latency between Fog nodes is higher. This is because consistency level \texttt{ONE} does not need coordination between multiple replica nodes, hence, avoiding additional round trips in the network. The coordinator that receives a query can directly execute it locally and return the result to the client.

Second, we see that already small changes in network latency have a high impact on the latencies in higher consistency levels. This is because of the required coordination between replicas, which implies additional network round trips. 

From the preliminary studies, we can make some important observations. First, the most beneficial consistency decisions in context-sensitive differential consistency are between consistency level \texttt{ONE} and any other high consistency level. Second, if high consistency levels must be enforced, it is particularly important to place the replicas close to each other. 

This fits well to the design goals of FogStore. First, the Fog-aware replica placement algorithm treats network latency as a first-class citizen, so that replicas will be placed as close as possible to each other and to the clients. Second, the context-sensitive differential consistency allows for fast responses when clients can deal with reduced consistency in their current context.

%% file: content/related.tex
\section{Related Work}



 Highly specialized state management protocols can allow for high throughput and low latency, e.g., in VNFs \cite{7502432, 7474417}. However, a customized state management for each single application is error-prone and hinders state sharing between multiple different applications. 
Instead, it is often more practicable to use existing general-purpose distributed data stores. This proposition is in line with a recent industry experience paper from Davie et al. \cite{Davie:2017:DAS:3041027.3041030}, where the authors achieve \emph{interoperability} between different SDN controllers by employing an external data base.

DDSs operate in a trade-off between availability, consistency, latency, and partitioning tolerance. The CAP theorem \cite{brewer2000towards} states that in times of network partitions, the data store can choose between data consistency and availability, while the PACELC theorem \cite{6127847} extends CAP, pointing out a trade-off between consistency and latency in times when the network is not partitioned. Regarding transactional consistency, the available data stores either provide strict ACID properties, or BASE (Basically Available, Soft state, Eventually consistent) \cite{Pritchett:2008:BAA:1394127.1394128} properties. 

Chun et al. \cite{Chun:2006:ERM:1267680.1267684} address the question of \emph{how many} replicas of a data record are needed to survive a given failure rate of disks by modeling disk failure and data replication as a birth-death process. The replication degree is an important question to be considered in FogStore as well; however, it is questionable whether failures in a Fog architecture can be modeled in the same way as disk failures in a data center. Doceur and Wattenhofer \cite{douceur2001competitive} place a given number of replicas on nodes such that the data availability is maximized; however, they do not take into account communication latency, which is a crucial factor in Fog Computing. The latency-aware replica placement algorithm proposed by Szymaniak et al. \cite{szymaniak2006latency} is based on the assumption that placing more replicas in a heavily-loaded network region improves the access latency for a large number of clients. This does not apply to Fog Computing scenarios where the locality of clients and data plays a prominent role.

There are approaches to adapt the consistency level of DDSs based on different performance metrics.
Harmony \cite{6337791} is a system that adapts consistency levels based on the probability of stale reads. The Bismar system \cite{6546112} employs monetary cost as a performance metric for adapting consistency levels. Aslan and Matrawy \cite{7751168} propose a framework that implements tunable consistency based on arbitrary performance goals of the application.
Different from FogStore, all of these systems do not take into account the context of data and clients and only provide a global consistency setting that treats all client requests as a black box.



%% file: content/conclusion.tex
\section{Conclusion}
Distributed data stores are an important building block for stateful applications and VNFs. Pushing the data stores to the Fog Computing continuum requires to devise new placement strategies. Further, the context-awareness of Fog Computing scenarios can be utilized in adapting the consistency level to the data and client context. To this end, we propose the FogStore system that provides a Fog-aware replica placement algorithm and context-sensitive differential consistency.

